\documentclass[prd,aps,showpacs,twocolumn,nofootinbib,preprintnumbers]{revtex4}
\usepackage{graphicx,color,amsmath,amsxtra}
\usepackage{epsf}
\usepackage{amssymb}
\usepackage{enumerate}
\usepackage{hhline}
\usepackage{array}
\usepackage{tabularx}

\usepackage[T1]{fontenc}
\usepackage[latin1]{inputenc}
\usepackage[english]{babel}
\usepackage{amsfonts}

\newcommand{\bear}{\begin{array}}  \newcommand{\eear}{\end{array}}
\newcommand{\bea}{\begin{eqnarray}}  \newcommand{\eea}{\end{eqnarray}}
\newcommand{\beq}{\begin{equation}}  \newcommand{\eeq}{\end{equation}}
\newcommand{\bef}{\begin{figure}}  \newcommand{\eef}{\end{figure}}
\newcommand{\bec}{\begin{center}}  \newcommand{\eec}{\end{center}}

\newcommand{\Eqn}[1]{&\hspace{-0.2em}#1\hspace{-0.2em}&}

\def\be{\begin{equation}}
\def\ee{\end{equation}}
\def\bea{\begin{eqnarray}}
\def\eea{\end{eqnarray}}
\def\beaa{\begin{eqnarray*}}
\def\eeaa{\end{eqnarray*}}
\def\beq{\begin{eqnarray}}
\def\eeq{\end{eqnarray}}

\def\nn{\nonumber \\}

\begin{document}

\title{Equivalence of modified gravity equation to the Clausius relation
}

\author{Kazuharu Bamba$^{1,}$, 
Chao-Qiang Geng$^{1,}$, 
Shin'ichi Nojiri$^{2,}$ 
and Sergei D. Odintsov$^{3,}$\footnote{Also at Tomsk State
Pedagogical University. }}
\affiliation{
$^1$Department of Physics, National Tsing Hua University, Hsinchu, Taiwan 300\\
$^2$Department of Physics, Nagoya University, Nagoya 464-8602, Japan\\
$^3$Instituci\`{o} Catalana de Recerca i Estudis Avan\c{c}ats (ICREA)
and Institut de Ciencies de l'Espai (IEEC-CSIC),
Campus UAB, Facultat de Ciencies, Torre C5-Par-2a pl, E-08193 Bellaterra
(Barcelona), Spain
}


\begin{abstract}

We explicitly show that the equations of motion for modified gravity theories 
of $F(R)$-gravity, the scalar-Gauss-Bonnet gravity, 
$F(\mathcal{G})$-gravity and the non-local gravity 
are equivalent to the Clausius relation in thermodynamics. 
In addition, we discuss the relation between 
the expression of the entropy and the contribution from the modified gravity
as well as the matter to the definition of the energy flux (heat).

\end{abstract}


\maketitle


The discovery of black hole (BH) entropy by
Bekenstein~\cite{Bekenstein:1973ur} and the first law of
BH thermodynamics~\cite{Bardeen:1973gs} with a Hawking
temperature~\cite{Hawking:1974sw} implies the fundamental connection
between gravitation and thermodynamics.
Jacobson has shown that the Einstein equation can be derived from the Clausius
relation,
$\delta S = \delta Q/T$, in thermodynamics~\cite{Jacobson:1995ab}.
Here, $S$ is the entropy, $Q$ is the heat, and $T$ is the temperature.
In his formulation, $\delta Q$ is interpreted as the energy flux through the
local Rindler horizon $\mathcal{H}$ at a free-falling local observer $p_0$:
\be
\label{BH00}
\delta Q = \int_{\mathcal{H}} T_{\mu\nu} \chi^{\mu} d \Sigma^{\nu}\,,
\ee
and $T$ could be the Unruh temperature~\cite{Unruh:1976db}:
\be
\label{BH01}
T=\frac{k}{2\pi}\,.
\ee
In Eq.~(\ref{BH00}), $\chi^\mu$ is an approximate local boost Killing field
future directed to the past of $p_0$, $T_{\mu\nu}$ is the contribution to the
energy-momentum tensor from all ordinary matters, and the integration is over
a pencil of generators of $\mathcal{H}$ at $p_0$,
and $d \Sigma^{\mu}$ is given by
$d \Sigma^{\mu} = K^{\mu} d\lambda dA$, where
$dA$ is the cross section area element of $\mathcal{H}$ and $K^\mu$ is an
approximate Killing field generating boost at $p_0$ and vanishing at $p_0$,
which is taken as the future pointing to the inside past of $p_0$.
In Eq.~(\ref{BH01}), $k$ is the acceleration of the Killing orbit on which the
norm of $\chi^{\mu}$ is unity
if $K$ is a tangent vector to the generators of $\mathcal{H}$ with
an affine parameter $\lambda$ ($\lambda = 0$ at $p_0$).
The formulation was extended to
$F(R)$-gravity~\cite{Elizalde:2008pv, Eling:2006aw}
(for a review of $F(R)$ and other modified gravity theories,
see~\cite{Nojiri:2006ri}),
and to the more general extended gravity theory~\cite{Brustein:2009hy} 
(see also related discussions in~\cite{Parikh:2009qs}). 
%
The first generalization of the relation between 
the gravitational field equations and thermodynamics to 
Lanczos-Lovelock gravity has been executed in~\cite{Padmanabhan:2006ag, 
Paranjape:2006ca} 
(see also a recent related work in~\cite{Kothawala:2009kc}). 
Moreover, 
the entropy functional approach for Lanczos-Lovelock gravity 
has extensively been discussed in~\cite{Padmanabhan:2007xy}. 
This entropy functional approach is the same (except for a four-divergence) as 
the one used in~\cite{Brustein:2009hy, Wu:2008rp} 
and its connection with diffeomorphism invariance has been noted 
in~\cite{Padmanabhan:2009ry}, in which 
it has been stated that in principle, 
all diffeomorphism invariant theories have an entropic 
derivation, provided one is willing to accept a particular expression as 
entropy. 
%

In~\cite{Elizalde:2008pv} and~\cite{Brustein:2009hy}, the definition of the
entropy $S$ was used as~\cite{Wald:1993nt}:
\be
\label{BH02}
S = - \frac{2}{T} \oint_{\partial\mathcal{H}}
S^{\mu\rho\nu\sigma} \hat\epsilon_{\mu\rho} \epsilon_{\nu\sigma}\,,
\ee
with
\be
\label{BH2}
S^{\mu\rho\nu\sigma} \equiv \frac{1}{\sqrt{-g}} \frac{\delta I}{\delta R_{\mu\rho\nu\sigma} }\,,
\ee
where $I$ is the action,
the integration in Eq.~(\ref{BH02}) is over the surface enclosing the
volume $\mathcal{H}$, $\epsilon^{\mu\nu}$ is a 2-dimensional volume form, and
$\hat\epsilon^{\mu\nu}$ is given by
$\hat\epsilon^{\mu\nu} = \nabla^\mu \tilde\chi^\nu
= \epsilon^{\mu\nu}/ \bar\epsilon$
with $\tilde\chi^\nu = \chi^\nu/k$ and 
the area element $\bar\epsilon$ of the cross section of the horizon.
Consequently, in~\cite{Brustein:2009hy} the following formula has been
obtained:\footnote{
The sign of the first term in the right-hand side (r.h.s.) is different from
that in~\cite{Brustein:2009hy},
which comes from the difference of the definition of the Riemann tensor.
In the present paper, we define
the curvatures as,
\[
R = g^{\mu\nu}R_{\mu\nu} \,, \quad
R_{\mu\nu}= R^\lambda_{\ \mu\lambda\nu} \,,
\]
\[
R^\lambda_{\ \mu\rho\nu} =
-\Gamma^\lambda_{\mu\rho,\nu}
+ \Gamma^\lambda_{\mu\nu,\rho}
- \Gamma^\eta_{\mu\rho}\Gamma^\lambda_{\nu\eta}
+ \Gamma^\eta_{\mu\nu}\Gamma^\lambda_{\rho\eta} \,.
\]
}
\be
\label{BH1}
T^{\sigma\nu} = 2 \left[ -2 \nabla_\mu \nabla_\rho S^{\mu\sigma\nu\rho} +
S^{\mu\rho\tau\sigma} R_{\mu\rho\tau}^{\ \ \ \nu} \right] + g^{\sigma\nu}
\Phi\,.
\ee
Here, $\Phi$ is determined by the conservation law or Bianchi identity.
As a result, the equivalence between the equations of motion and the
fundamental thermodynamic relation for the generalized theories of gravity
has been demonstrated in~\cite{Brustein:2009hy}.

In the present paper, we study modified gravity theories, 
in particular 
(i) $F(R)$-gravity\footnote{
The equations of motion for $F(R)$-gravity in~\cite{Eling:2006aw} were
derived by using nonequilibrium thermodynamics, while
in~\cite{Elizalde:2008pv} they were examined in
equilibrium thermodynamics with the idea of
``local-boost-invariance''~\cite{Iyer:1994ys}.
In the present paper, however, we apply the generalization of the Jacobson's
derivation proposed in~\cite{Brustein:2009hy} to $F(R)$-gravity.
},
(ii) the scalar-Gauss-Bonnet gravity inspired by (super)string theories,
(iii) $F(\mathcal{G})$-gravity~\cite{Nojiri:2005jg}
and
(iv) the non-local gravity,
where
$\mathcal{G} \equiv R^2 -4R_{\mu\nu}R^{\mu\nu} +
R_{\mu\nu\rho\sigma}R^{\mu\nu\rho\sigma}$ is the Gauss-Bonnet invariant 
($R$, $R_{\mu\nu}$ and $R_{\mu\nu\rho\sigma}$ are the scalar curvature, 
the Ricci tensor and the Riemann tensor, respectively) and
$F(\mathcal{G})$ is an appropriate function in terms of $\mathcal{G}$, 
and explicitly show that the equations of motion for these theories are 
equivalent to the Clausius relation in thermodynamics by applying the 
formula in Eq.~(\ref{BH1}). 
Extended (or modified) gravity theories are frequently studied in the 
context of effective gravity theories of string theories and supergravity.
In addition, these theories have the capability to explain
the current accelerated expansion of the universe
alternative to the $\Lambda$CDM cosmology
as well as inflation in the early universe.

In Eq.~(\ref{BH00}),
the energy flux $\delta Q$ is expressed by using
the energy-momentum tensor, $T_{\mu\nu}$, of all ordinary matters.
The entropy $S$ is defined by Eq.~(\ref{BH02}) with Eq.~(\ref{BH2}) for a
gravity theory.
$\delta Q$ is related to $S$ through the Clausius relation
$\delta S = \delta Q/T$.
By applying Eqs.~(\ref{BH00}) and (\ref{BH02}) to the Clausius relation,
Eq.~(\ref{BH1}) can be derived.
Hence, Eq.~(\ref{BH1}) corresponds to the equation of motion for the
gravity theory, which relates the matter and gravity.
For the above four extended gravity theories, we derive the explicit
expressions of $\nabla_\mu \nabla_\rho S^{\mu\sigma\nu\rho}$,
$S^{\mu\rho\tau\sigma} R_{\mu\rho\tau}^{\ \ \ \nu}$
and $\Phi$, which are components on the r.h.s. of Eq.~(\ref{BH1}),
to clearly illustrate the equations of motion from
the Clausius relation in thermodynamics.
Our investigation is the application of the Jacobson's proposal to derive
the Einstein equation as a thermodynamic equation of state
in general relativity to modified gravity theories.
We use units of $k_\mathrm{B} = c = \hbar = 1$ and denote the
gravitational constant $8 \pi G$ by
${\kappa}^2 \equiv 8\pi/{M_{\mathrm{Pl}}}^2$
with the Planck mass of $M_{\mathrm{Pl}} = G^{-1/2} = 1.2 \times 10^{19}$GeV.


A general form of the action describing modified gravity theories (in the
below context) is given by
\begin{eqnarray}
I = \int d^4 x \sqrt{-g} \left[ \frac{\mathcal{F}(R,\phi,X,\mathcal{G})}
{2\kappa^2} +
{\mathcal{L}}_{\mathrm{matter}} \right]\,,
\label{eq:2.1}
\end{eqnarray}
where $g$ is the determinant of the metric tensor $g_{\mu\nu}$,
${\mathcal{L}}_{\mathrm{matter}}$ is the matter Lagrangian,
$\phi$ is a scalar field,
$X \equiv -\left(1/2\right) g^{\mu\nu} {\nabla}_{\mu}\phi {\nabla}_{\nu}\phi$
is a kinetic term of $\phi$ (${\nabla}_{\mu}$ is the covariant derivative
operator associated with $g_{\mu \nu}$), and
$\mathcal{F}(R,\phi,X,\mathcal{G})$ is an arbitrary function in terms of
$R$, $\phi$, $X$ and $\mathcal{G}$.
The scalar field $\phi$ is a gravitational partner, e.g., a dilaton,
in the case of string theories.

 From the action in Eq.~(\ref{eq:2.1}), the gravitational field equation and
the equation of motion for $\phi$ are derived as
\begin{eqnarray}
&& \hspace{-5mm}
{\mathcal{F}}_{,R} \left( R_{\mu\nu}-\frac{1}{2}R g_{\mu\nu}\right)
= \kappa^2 T^{(\mathrm{matter})}_{\mu \nu}
+\frac{1}{2}g_{\mu\nu} \left(\mathcal{F}-{\mathcal{F}}_{,R}R\right)
\nonumber \\
&& \hspace{-5mm}
+{\nabla}_{\mu}{\nabla}_{\nu}
{\mathcal{F}}_{,R} -g_{\mu\nu} \Box {\mathcal{F}}_{,R}
+\frac{1}{2} {\mathcal{F}}_{,X} \partial_{\mu} \phi \partial_{\nu} \phi
\nonumber \\
&& \hspace{-5mm}
+\bigl(-2RR_{\mu\nu} +4R_{\mu\rho}R_{\nu}{}^{\rho}
-2R_{\mu}{}^{\rho\sigma\tau}R_{\nu\rho\sigma\tau}
\nonumber \\
&& \hspace{-5mm}
+4g^{\alpha\rho}g^{\beta\sigma}R_{\mu\alpha\nu\beta}R_{\rho\sigma}
\bigr){\mathcal{F}}_{,\mathcal{G}}
+2\left({\nabla}_{\mu}{\nabla}_{\nu}{\mathcal{F}}_{,\mathcal{G}}\right)R
\nonumber \\
&& \hspace{-5mm}
-2g_{\mu \nu}\left(\Box {\mathcal{F}}_{,\mathcal{G}}\right)R
+4\left(\Box {\mathcal{F}}_{,\mathcal{G}}\right)R_{\mu \nu}
-4\left({\nabla}_{\rho}{\nabla}_{\mu}{\mathcal{F}}_{,\mathcal{G}}\right)
R_{\nu}{}^{\rho}
\nonumber \\
&& \hspace{-5mm}
-4\left({\nabla}_{\rho}{\nabla}_{\nu}{\mathcal{F}}_{,\mathcal{G}}\right)
R_{\mu}{}^{\rho}
+4g_{\mu \nu}\left({\nabla}_{\rho}{\nabla}_{\sigma}
{\mathcal{F}}_{,\mathcal{G}}\right)R^{\rho\sigma}
\nonumber \\
&& \hspace{-5mm}
-4\left({\nabla}_{\rho}{\nabla}_{\sigma}{\mathcal{F}}_{,\mathcal{G}}\right)
g^{\alpha\rho}g^{\beta\sigma}R_{\mu\alpha\nu\beta}\,,
\label{eq:2.2} \\
&& \hspace{-5mm}
{\mathcal{F}}_{,\phi} +\frac{1}{\sqrt{-g}}\partial_\mu
\left({\mathcal{F}}_{,X} \sqrt{-g} g^{\mu\nu}
\partial_\nu \phi\right) = 0\,,
\label{eq:2.3}
\end{eqnarray}
where we have used the following expressions:
\begin{eqnarray}
\hspace{-5mm}
&&
{\mathcal{F}}_{,R} =
\frac{\partial \mathcal{F}(R,\phi,X,\mathcal{G})}{\partial R}\,, \quad
{\mathcal{F}}_{,X} =
\frac{\partial \mathcal{F}(R,\phi,X,\mathcal{G})}{\partial X}\,,
\nonumber \\
\hspace{-5mm}
&&
{\mathcal{F}}_{,\mathcal{G}} =
\frac{\partial \mathcal{F}(R,\phi,X,\mathcal{G})}
{\partial \mathcal{G}}\,, \quad
{\mathcal{F}}_{,\phi} =
\frac{\partial \mathcal{F}(R,\phi,X,\mathcal{G})}{\partial \phi}\,.
\label{eq:2.4}
\end{eqnarray}
Here, $\Box \equiv g^{\mu \nu} {\nabla}_{\mu} {\nabla}_{\nu}$
is the covariant d'Alembertian for a scalar field and
$T^{(\mathrm{matter})}_{\mu \nu}$
is the contribution to the energy-momentum tensor from all ordinary matters.

Hereafter, we investigate four explicit examples of modified gravity theories.

\vspace{2mm}
\noindent
{\bf (i) $F(R)$-{\em gravity}}
\vspace{1mm}

 From the action
\be
I = \int d^4 x \sqrt{-g} \left[ F(R)
+ {\mathcal{L}}_{\mathrm{matter}} \right]\,,
\label{FR1}
\ee
we obtain
\be
\label{FR2}
S^{\mu\nu\rho\sigma} = \frac{F'(R)}{2}\left(g^{\mu\nu} g^{\rho\sigma}
 - g^{\mu\sigma} g^{\nu\rho} \right) \,,
\ee
and
\begin{eqnarray}
\label{FR3-1}
\nabla_\mu \nabla_\sigma S^{\mu\rho\nu\sigma}
\Eqn{=} \frac{1}{2} \left(\nabla^\nu \nabla^\rho - g^{\nu\rho} \Box \right)
F'(R)\,,
\\
\label{FR3-2}
S^{\mu\rho\tau\sigma} R_{\mu\rho\tau}^{\ \ \ \ \nu} \Eqn{=}
R^{\sigma\nu} F'(R)\,.
\end{eqnarray}
Here and in what follows, the prime denotes differentiation with respect to
the argument of the function $F$ as $F'(R) = dF(R)/dR$.
On the other hand, the equation in $F(R)$-gravity, corresponding to the
Einstein equation, is given by
\begin{eqnarray}
\label{M2}
0 \Eqn{=}
\frac{1}{2}g_{\mu\nu} F(R) - R_{\mu\nu}F'(R) + \nabla_\mu \nabla_\nu F'(R)
\nonumber \\
&& \hspace{0mm}
 - g_{\mu\nu}\Box F'(R)
+ \frac{1}{2} T^{(\mathrm{matter})}_{\mu\nu}\,.
\end{eqnarray}
By comparing Eq.~(\ref{M2}) with Eq.~(\ref{BH1}) and using Eqs.~(\ref{FR3-1})
and (\ref{FR3-2}), we find
\begin{eqnarray}
\label{B-1}
T_{\mu\nu} \Eqn{=} T^{(\mathrm{matter})}_{\mu \nu}\,,
\\
\label{FR4}
\Phi \Eqn{=} - F(R)\,.
\end{eqnarray}

\vspace{2mm}
\noindent
{\bf (ii) {\em Scalar-Gauss-Bonnet gravity}}
\vspace{1mm}

The action is given by
\begin{eqnarray}
I \Eqn{=} \int d^4 x \sqrt{-g} \biggl[ \frac{R}{2\kappa^2}
-\frac{\gamma}{2} g^{\mu\nu} {\partial}_{\mu}\phi {\partial}_{\nu}\phi
-V(\phi)+ f(\phi)\mathcal{G}
\nonumber \\
&& \hspace{20mm}
+ {\mathcal{L}}_{\mathrm{matter}} \biggr]\,,
\label{eq:1}
\end{eqnarray}
where $\gamma = \pm1$.
If $\phi$ is a canonical scalar field, $\gamma =1$.
On the other hand,
if the GB invariant is not included, $\phi$ behaves as a phantom scalar field
only when $\gamma = -1$.
Moreover, $V(\phi)$ is the potential and
$f(\phi)$ is an appropriate function of $\phi$.
For this theory, we find
\bea
\label{BH3}
S^{\mu\rho\nu\sigma} \Eqn{=}
\frac{1}{4\kappa^2}\left(g^{\mu\nu} g^{\rho\sigma} - g^{\mu\sigma} g^{\nu\rho}
\right)
\nonumber \\
&& \hspace{0mm}
+ f(\phi) \bigl\{ \left(g^{\mu\nu} g^{\rho\sigma} - g^{\mu\sigma} g^{\nu\rho}
\right) R
\nonumber \\
&& \hspace{0mm}
 - 2 \left(g^{\rho\sigma} R^{\mu\nu} - g^{\rho\nu} R^{\mu\sigma} - g^{\mu\sigma} R^{\rho\nu}
+ g^{\mu\nu} R^{\rho\sigma} \right)
\nonumber \\
&& \hspace{0mm}
+ 2 R ^{\mu\rho\nu\sigma}\bigr\}\,,
\eea
and
\bea
\label{BH4}
&& \nabla_\sigma S^{\mu\rho\nu\sigma}
= \nabla_\sigma f(\phi) \bigl\{ \left(g^{\mu\nu} g^{\rho\sigma} - g^{\mu\sigma} g^{\nu\rho} \right) R
\nonumber \\
&& \hspace{5mm}
 - 2 \left(g^{\rho\sigma} R^{\mu\nu} - g^{\rho\nu} R^{\mu\sigma} - g^{\mu\sigma} R^{\rho\nu} \right)
+ 2 R ^{\mu\rho\nu\sigma}\bigr\}
\nn
&& \hspace{5mm}
+ f(\phi) \bigl\{ \left(g^{\mu\nu} g^{\rho\sigma} - g^{\mu\sigma} g^{\nu\rho} \right) \nabla_\sigma R
\nonumber \\
&& \hspace{5mm}
 - 2 \left(\nabla^\rho R^{\mu\nu} - g^{\rho\nu} \nabla_\sigma R^{\mu\sigma} - \nabla^\mu R^{\rho\nu}
+ g^{\mu\nu} \nabla_\sigma R^{\rho\sigma} \right)
\nonumber \\
&& \hspace{5mm}
+ 2 \nabla_\sigma R ^{\mu\rho\nu\sigma} \bigr\}\,.
\eea
 From the Bianchi identity:
\be
\label{Bianchi}
0=\nabla_\mu R_{\nu\rho\sigma\tau} + \nabla_\nu R_{\rho\mu\sigma\tau}
+ \nabla_\rho R_{\mu\nu\sigma\tau}\,,
\ee
we have several identities:
\be
\label{GBii}
\nabla^\rho R_{\rho\tau\mu\nu} = \nabla_\mu R_{\nu\tau} - \nabla_\nu R_{\mu\tau}\ ,\quad
\nabla^\rho R_{\rho\mu} = \frac{1}{2} \nabla_\mu R\,.
\ee
We see that the terms multiplied by $f(\phi)$ (without $\nabla_\sigma$) cancel
with each other and we get
\begin{eqnarray}
\label{BH5}
&& \hspace{-8mm}
\nabla_\sigma S^{\mu\rho\nu\sigma}
= \nabla_\sigma f(\phi) \bigl\{ \left(g^{\mu\nu} g^{\rho\sigma} - g^{\mu\sigma} g^{\nu\rho} \right) R
\nonumber \\
&& \hspace{-3mm}
 - 2 \left(g^{\rho\sigma} R^{\mu\nu} - g^{\rho\nu} R^{\mu\sigma} - g^{\mu\sigma} R^{\rho\nu} \right)
+ 2 R ^{\mu\rho\nu\sigma}\bigr\}\,.
\end{eqnarray}
Similarly, we find
\begin{eqnarray}
\label{BH6}
&& \hspace{-8mm}
\nabla_\mu \nabla_\sigma S^{\mu\rho\nu\sigma}
= \nabla_\mu \nabla_\sigma f(\phi) \bigl\{ \left(g^{\mu\nu} g^{\rho\sigma}
 - g^{\mu\sigma} g^{\nu\rho} \right) R
\nonumber \\
&& \hspace{-3mm}
 - 2 \left(g^{\rho\sigma} R^{\mu\nu} - g^{\rho\nu} R^{\mu\sigma} - g^{\mu\sigma} R^{\rho\nu} \right)
+ 2 R ^{\mu\rho\nu\sigma}\bigr\} \,,
\end{eqnarray}
and
\begin{eqnarray}
\label{BH7}
S^{\mu\rho\nu\sigma}R_{\mu\rho\nu}^{\ \ \ \eta}
\Eqn{=}
\frac{1}{2\kappa^2} R^{\sigma\eta} + f(\phi) \bigl\{ 2R^{\sigma\eta} R
 - 4 R_{\mu\ \nu}^{\ \sigma\ \eta} R^{\mu\nu}
\nonumber \\
&& \hspace{0mm}
-4 R_\mu^{\ \eta} R^{\mu\sigma}
+ 2 R_{\mu\rho\nu}^{\ \ \ \eta} R^{\mu\rho\nu\sigma} \bigr\}\,.
\end{eqnarray}
In four dimensions, we have the following non-trivial identity:
\begin{eqnarray}
\label{4d_GBidentity}
0 \Eqn{=}
\frac{1}{2}g_{\mu\nu} \mathcal{G} -2RR_{\mu\nu} +4R_{\mu\rho}R_{\nu}{}^{\rho}
-2R_{\mu}{}^{\rho\sigma\tau}R_{\nu\rho\sigma\tau}
\nonumber \\
&& \hspace{0mm}
+4g^{\alpha\rho}g^{\beta\sigma}R_{\mu\alpha\nu\beta}R_{\rho\sigma}\,.
\end{eqnarray}
We can rewrite Eq.~(\ref{BH7}) as
\be
\label{BH8}
S^{\mu\rho\nu\sigma}R_{\mu\rho\nu}^{\ \ \ \eta}
= \frac{1}{2\kappa^2} R^{\sigma\eta} + \frac{f(\phi) \mathcal{G}}{2} g^{\sigma\eta}\,.
\ee
Now by comparing Eq.~(\ref{BH1}) with the equation in the scalar-Gauss-Bonnet
gravity corresponding to the Einstein equation:
\bea
&&
T^{(\mathrm{matter})}_{\mu \nu} =
\frac{1}{\kappa^2}\left(R_{\mu\nu} - \frac{1}{2}R g_{\mu\nu}\right)
\nonumber \\
&& \hspace{0mm}
 - \gamma \left( \partial_{\mu} \phi \partial_{\nu} \phi
 - \frac{1}{2} g_{\mu\nu} \partial_{\rho} \phi \partial^{\rho} \phi \right)
+ g_{\mu\nu} V(\phi)
\nn
&& \hspace{0mm}
- 4 \left({\nabla}_{\mu}{\nabla}_{\nu}f(\phi)\right)R
 + 4 g_{\mu \nu}\left(\Box f(\phi)\right)R
 - 8 \left(\Box f(\phi)\right)R_{\mu \nu}
\nonumber \\
&& \hspace{0mm}
 + 8 \left({\nabla}_{\rho}{\nabla}_{\mu} f(\phi)\right)R_{\nu}{}^{\rho}
+ 8 \left({\nabla}_{\rho}{\nabla}_{\nu} f(\phi)\right)R_{\mu}{}^{\rho}
\nonumber \\
&& \hspace{0mm}
 -  8 g_{\mu \nu}\left({\nabla}_{\rho}{\nabla}_{\sigma} f(\phi)\right) R^{\rho\sigma}
\nonumber \\
&& \hspace{0mm}
 + 8 \left({\nabla}_{\rho}{\nabla}_{\sigma} f(\phi)\right)
 g^{\alpha\rho}g^{\beta\sigma}R_{\mu\alpha\nu\beta}\,,
\label{eq:2}
\eea
we get
\begin{eqnarray}
\label{BH9-1}
T_{\mu \nu} \Eqn{=}
T^{(\mathrm{matter})}_{\mu \nu} + \gamma \left( \partial_{\mu}
\phi \partial_{\nu} \phi
 - \frac{1}{2} g_{\mu\nu} \partial_{\rho} \phi \partial^{\rho} \phi \right)
\nonumber \\
&& \hspace{0mm}
 - g_{\mu\nu} V(\phi)\,,
\\
\label{BH9-2}
\Phi \Eqn{=} \frac{R}{2\kappa^2} - f(\phi)\mathcal{G}\,.
\end{eqnarray}

\begin{table*}[tbp]
\caption{Explicit expressions of
$\nabla_\mu \nabla_\rho S^{\mu\sigma\nu\rho}$,
$S^{\mu\rho\tau\sigma} R_{\mu\rho\tau}^{\ \ \ \nu}$
and $\Phi$ in Eq.~(\ref{BH1}) for four modified gravity theories:
(i) $F(R)$-gravity, (ii) the scalar-Gauss-Bonnet gravity,
(iii) $F(\mathcal{G})$-gravity
and (iv) the non-local gravity.
}
\begin{center}
\begin{tabular}
{llll}
\hline
\hline
Theory
& $\nabla_\mu \nabla_\rho S^{\mu\sigma\nu\rho}$
& $S^{\mu\rho\tau\sigma} R_{\mu\rho\tau}^{\ \ \ \nu}$
& $\Phi$
\\[0mm]
\hline
(i)
$F(R)$-gravity
&
$\frac{1}{2} \left(\nabla^\nu \nabla^\sigma - g^{\nu\sigma} \Box \right)
F'(R)$
& $R^{\sigma\nu} F'(R)$
& $- F(R)$
\\[0mm]
(ii)
Scalar-Gauss-Bonnet gravity
&
$\nabla_\mu \nabla_\rho f(\phi) \bigl\{ \left(g^{\mu\nu} g^{\sigma\rho}
 - g^{\mu\rho} g^{\nu\sigma} \right) R$
& $\frac{1}{2\kappa^2} R^{\sigma\nu} + \frac{f(\phi) \mathcal{G}}{2}
g^{\sigma\nu}$
& $\frac{R}{2\kappa^2} - f(\phi)\mathcal{G}$
\\[0mm]
&$ - 2 \left(g^{\sigma\rho} R^{\mu\nu} - g^{\sigma\nu} R^{\mu\rho}
- g^{\mu\rho} R^{\sigma\nu} \right)$
&
&
\\[0mm]
&$+ 2 R ^{\mu\sigma\nu\rho}\bigr\}$
&
&
\\[0mm]
(iii)
$F(\mathcal{G})$-gravity
& $\nabla_\mu \nabla_\rho F'(\mathcal{G}) \bigl\{ \left(g^{\mu\nu}
g^{\sigma\rho}
 - g^{\mu\rho} g^{\nu\sigma} \right) R$
& $\frac{1}{2\kappa^2} R^{\sigma\nu} +
\frac{F'(\mathcal{G}) \mathcal{G}}{2} g^{\sigma\nu}$
& $\frac{R}{2\kappa^2} - F(\mathcal{G})$
\\[0mm]
&
$ - 2 \left(g^{\sigma\rho} R^{\mu\nu} - g^{\sigma\nu} R^{\mu\rho}
- g^{\mu\rho} R^{\sigma\nu} \right)$
&
&
\\[0mm]
& $+ 2 R ^{\mu\sigma\nu\rho}\bigr\}$
&
&
\\[0mm]
(iv)
Non-local gravity
& $\frac{1}{4\kappa^2}\left(\nabla^\nu \nabla^\sigma - g^{\nu\sigma}
\Box \right) \left(\tilde{f}(\varphi) -\xi\right)$
& $\frac{1}{2\kappa^2}\left( 1 + \tilde{f}(\varphi) - \xi\right)R^{\sigma\nu}$
& $\frac{R}{2\kappa^2}\left( 1 + \tilde{f}(\varphi) - \xi\right)$
\\[1mm]
\hline
\hline
\end{tabular}
\end{center}
\label{tb:table1}
\end{table*}

\vspace{2mm}
\noindent
{\bf (iii) $F(\mathcal{G})$-\noindent{\em gravity}}
\vspace{1mm}

In the so-called $F(\mathcal{G})$-gravity~\cite{Nojiri:2005jg}, the action is
given by
\be
I = \int d^4 x \sqrt{-g} \left[ \frac{R}{2\kappa^2}
+ F(\mathcal{G}) + {\mathcal{L}}_{\mathrm{matter}} \right]\,.
\label{BH10}
\ee
In this case, we find
\bea
\label{BH11}
S^{\mu\rho\nu\sigma}
\Eqn{=}
\frac{1}{4\kappa^2}\left(g^{\mu\nu} g^{\rho\sigma} - g^{\mu\sigma} g^{\nu\rho}
\right)
\nn
&&
+ F'(\mathcal{G}) \bigl\{ \left(g^{\mu\nu} g^{\rho\sigma} - g^{\mu\sigma}
g^{\nu\rho} \right) R
\nonumber \\
&& \hspace{0mm}
 - 2 \left(g^{\rho\sigma} R^{\mu\nu} - g^{\rho\nu} R^{\mu\sigma} - g^{\mu\sigma} R^{\rho\nu}
+ g^{\mu\nu} R^{\rho\sigma} \right)
\nonumber \\
&& \hspace{0mm}
+ 2 R ^{\mu\rho\nu\sigma}\bigr\}\,,
\eea
where $F'(\mathcal{G}) = dF(\mathcal{G})/d\mathcal{G}$.
By repeating the calculations similar to Eqs.~(\ref{BH4})--(\ref{BH8}),
we obtain
\bea
\label{BH12}
\nabla_\mu \nabla_\sigma S^{\mu\rho\nu\sigma} \Eqn{=}
\nabla_\mu \nabla_\sigma F'(\mathcal{G}) \bigl\{ \left(g^{\mu\nu}
g^{\rho\sigma}
 - g^{\mu\sigma} g^{\nu\rho} \right) R
\nonumber \\
&& \hspace{-21mm}
 - 2 \left(g^{\rho\sigma} R^{\mu\nu} - g^{\rho\nu} R^{\mu\sigma} - g^{\mu\sigma} R^{\rho\nu} \right)
+ 2 R ^{\mu\rho\nu\sigma}\bigr\} \,, \\
\label{BH13}
S^{\mu\rho\nu\sigma}R_{\mu\rho\nu}^{\ \ \ \eta}
\Eqn{=} \frac{1}{2\kappa^2} R^{\sigma\eta} +
\frac{F'(\mathcal{G}) \mathcal{G}}{2} g^{\sigma\eta}\,.
\eea
The equation of motion corresponding to the Einstein equation is given by
\begin{eqnarray}
T^{(\mathrm{matter})}_{\mu\nu} \Eqn{=}
\frac{1}{\kappa^2}\left(R_{\mu\nu} - \frac{1}{2}R g_{\mu\nu}\right)
 - g_{\mu\nu}\left( F(\mathcal{G}) - \mathcal{G} F'(\mathcal{G}) \right)
\nn
&& \hspace{-10mm}
- 4 \left({\nabla}_{\mu}{\nabla}_{\nu}F'(\mathcal{G})\right)R
+ 4 g_{\mu \nu}\left(\Box F'(\mathcal{G}) \right)R
\nonumber \\
&& \hspace{-10mm}
 - 8 \left(\Box F'(\mathcal{G}) \right)R_{\mu \nu}
+ 8 \left({\nabla}_{\rho}{\nabla}_{\mu} F'(\mathcal{G})
\right)R_{\nu}{}^{\rho}
\nonumber \\
&& \hspace{-10mm}
+ 8 \left({\nabla}_{\rho}{\nabla}_{\nu} F'(\mathcal{G}) \right)R_{\mu}{}^{\rho}
 - 8 g_{\mu \nu}\left({\nabla}_{\rho}{\nabla}_{\sigma} F'(\mathcal{G})
\right)R^{\rho\sigma}
\nonumber \\
&& \hspace{-10mm}
+ 8 \left({\nabla}_{\rho}{\nabla}_{\sigma} F'(\mathcal{G}) \right)
g^{\alpha\rho}g^{\beta\sigma}R_{\mu\alpha\nu\beta} \,.
\label{eq:26}
\end{eqnarray}
By using Eqs.~(\ref{BH12}) and (\ref{BH13}) and comparing Eq.~(\ref{BH1})
with Eq.~(\ref{eq:26}),
we find
\begin{eqnarray}
\label{BH14-1}
T_{\mu\nu} \Eqn{=} T^{(\mathrm{matter})}_{\mu \nu}\,,
\\
\label{BH14-2}
\Phi \Eqn{=} \frac{R}{2\kappa^2} - F(\mathcal{G})\,.
\end{eqnarray}

\vspace{2mm}
\noindent
{\bf (iv) {\em Non-local gravity}}
\vspace{1mm}

We now consider the non-local gravity~\cite{Deser:2007jk, Nojiri:2007uq}
\be
\label{nl1}
I=\int d^4 x \sqrt{-g}\left\{
\frac{1}{2\kappa^2}R\left(1 + \tilde{f}(\Box^{-1}R)\right) +
{\cal L}_\mathrm{matter}
\right\}\, .
\ee
Here, $\tilde{f}$ is an appropriate function in terms of its argument.
The above action can be rewritten by introducing two scalar fields $\varphi$
and $\xi$ in the following form~\cite{Nojiri:2007uq}:
\begin{eqnarray}
\label{nl2}
I \Eqn{=} \int d^4 x \sqrt{-g}\biggl[
\frac{1}{2\kappa^2}\left\{R\left(1 + \tilde{f}(\varphi)\right)
 - \partial_\mu \xi \partial^\mu \varphi - \xi R \right\}
\nonumber \\
&& \hspace{20mm}
+ {\cal L}_\mathrm{matter}
\biggr]\,,
\end{eqnarray}
which leads to
\be
\label{nl_S1}
S^{\mu\rho\nu\sigma}
= \frac{1}{4\kappa^2}\left( 1 + \tilde{f}(\varphi) - \xi \right)
\left(g^{\mu\nu} g^{\rho\sigma} - g^{\mu\sigma} g^{\nu\rho} \right)\,,
\ee
and
\bea
\label{nl_S2}
\hspace{-5mm}
\nabla_\mu \nabla_\sigma S^{\mu\rho\nu\sigma}
\Eqn{=}
\frac{1}{4\kappa^2}\left(\nabla^\nu \nabla^\rho - g^{\nu\rho} \Box \right)
\left(\tilde{f}(\varphi) -\xi\right)
\,, \\
\label{nl_S3}
\hspace{-5mm}
S^{\mu\rho\nu\sigma}R_{\mu\rho\nu}^{\ \ \ \eta}
\Eqn{=} \frac{1}{2\kappa^2}\left( 1 + \tilde{f}(\varphi) - \xi\right)
R^{\sigma\eta}\,.
\eea
The equation of motion corresponding to the Einstein equation is given by
\bea
\label{nl_S4}
&& \hspace{-5mm}
T^{(\mathrm{matter})}_{\mu\nu} = \frac{1}{\kappa^2} \biggl[
 - \frac{1}{2}g_{\mu\nu} \left\{ R\left(1 + \tilde{f}(\varphi) - \xi\right)
 - \partial_\rho \xi \partial^\rho \varphi \right\}
\nonumber \\
&& \hspace{-5mm}
+ R_{\mu\nu}\left(1 + \tilde{f}(\varphi) - \xi\right)
- \frac{1}{2}\left(\partial_\mu \xi \partial_\nu \varphi
+ \partial_\mu \varphi \partial_\nu \xi \right)
\nonumber \\
&& \hspace{-5mm}
+ \left(g_{\mu\nu}\Box - \nabla_\mu \nabla_\nu\right)\left(
\tilde{f}(\varphi) - \xi\right) \biggr]\,.
\eea
By using Eqs.~(\ref{nl_S2}) and (\ref{nl_S3}) and comparing Eq.~(\ref{BH1})
with Eq.~(\ref{nl_S4}),
we find
\begin{eqnarray}
\label{nl_S5}
T_{\mu\nu} \Eqn{=} T^{(\mathrm{matter})}_{\mu \nu}
+ \frac{1}{2\kappa^2}\bigl(\partial_\mu \xi \partial_\nu \varphi
+ \partial_\mu \varphi \partial_\nu \xi
\nonumber \\
&& \hspace{0mm}
- g_{\mu\nu} \partial_\rho \xi
\partial^\rho \varphi \bigr)\,,
\\
\label{nl_S6}
\Phi \Eqn{=} \frac{R}{2\kappa^2}\left( 1 + \tilde{f}(\varphi) -
\xi\right)\,.
\end{eqnarray}
We should note that there is an ambiguity in the separation 
into $T_{\mu\nu}$ part and $\Phi$ part. For example, instead of 
Eqs.~(\ref{nl_S5}) and (\ref{nl_S6}), 
we may choose 
\begin{eqnarray}
\label{nl_S5_B}
\tilde T_{\mu\nu} \Eqn{=} T^{(\mathrm{matter})}_{\mu \nu}
+ \frac{1}{2\kappa^2}\left(\partial_\mu \xi \partial_\nu \varphi
+ \partial_\mu \varphi \partial_\nu \xi \right) \,,
\\
\label{nl_S6_B}
\tilde \Phi \Eqn{=} \frac{R}{2\kappa^2}\left( 1 + \tilde{f}(\varphi) -
\xi\right) - \partial_\rho \xi \partial^\rho \varphi \,.
\end{eqnarray}
Here, the last term in Eq.~(\ref{nl_S5}) has been included in the definition 
of $\tilde\Phi$. 

 From the analysis of the above four modified gravity theories,
it is clear that we have derived explicit expressions of
$\nabla_\mu \nabla_\rho S^{\mu\sigma\nu\rho}$,
$S^{\mu\rho\tau\sigma} R_{\mu\rho\tau}^{\ \ \ \nu}$
and $\Phi$ in Eq.~(\ref{BH1}).
The results are summarized in Table~\ref{tb:table1}. 
A general expression for $\Phi$ can be expressed as 
the linear combination of $R/\left(2\kappa^2\right)$, which is the 
Lagrangian describing general relativity, and 
the Lagrangian of gravity ${\mathcal{L}}_{\mathrm{gravity}}$ as 
\begin{eqnarray}
\Phi = c_1 \frac{R}{2\kappa^2} + c_2 {\mathcal{L}}_{\mathrm{gravity}}\,, 
\label{eq:G-Phi}
\end{eqnarray}
where $c_1$ and $c_2$ are constants. 
For (i) $F(R)$-gravity, $c_1 = 0$, $c_2 = -1$ and 
${\mathcal{L}}_{\mathrm{gravity}} = F(R)$. 
For (ii) the scalar-Gauss-Bonnet gravity, 
$c_1 = 2$, $c_2 = -1$ and 
${\mathcal{L}}_{\mathrm{gravity}} = R/\left(2\kappa^2\right) + 
f(\phi)\mathcal{G}$. 
For (iii) $F(\mathcal{G})$-gravity, 
$c_1 = 2$, $c_2 = -1$ and 
${\mathcal{L}}_{\mathrm{gravity}} = 
R/\left(2\kappa^2\right) + F(\mathcal{G})$. 
For (iv) the non-local gravity, if we use the expression in Eq.~(\ref{nl_S6}), 
we find $c_1 = 0$, $c_2 = 1$ and 
${\mathcal{L}}_{\mathrm{gravity}} = 
\left[ R/\left(2\kappa^2\right) \right] 
\left(1 + \tilde{f}(\varphi) - \xi \right)$. 
If we use Eq.~(\ref{nl_S6_B}) instead of Eq.~(\ref{nl_S6}), we obtain 
${\mathcal{L}}_{\mathrm{gravity}} = 
\left[ R/\left(2\kappa^2\right) \right] 
\left(1 + \tilde{f}(\varphi) - \xi \right) - 
\partial_\rho \xi \partial^\rho \varphi $, 
which is nothing but the gravity part of the action in Eq.~(\ref{nl2}). 

Thus, we have reinforced the generalization~\cite{Brustein:2009hy} in 
modified gravity theories of the Jacobson's proposal to express
the Einstein equation as a thermodynamic equation of state
in general relativity with our analysis.
Our results could support the idea that gravitation on a macroscopic scale
is a manifestation of the thermodynamics of the vacuum state of quantum
field theory~\cite{Brustein:2009hy}. 

Between Eqs.~(\ref{nl_S5}-\ref{nl_S6}) and Eqs.~(\ref{nl_S5_B}-\ref{nl_S6_B}), 
we have find the ambiguity to define $T_{\mu\nu}$. As we see now, this could 
be a result from the ambiguity 
when we consider the thermodynamics in the extended gravities. 
In general, any gravity equation can be written as
\be
\label{BH15}
T^{(\mathrm{matter})}_{\mu\nu} + T^{(\mathrm{modified\ gravity})}_{\mu\nu}
= \frac{1}{\kappa^2}\left(R_{\mu\nu} - \frac{1}{2}R g_{\mu\nu}\right)\,.
\ee
Hence, if we include the contribution from
$T^{(\mathrm{modified\ gravity})}_{\mu\nu}$, which comes
from the modification of the Einstein gravity, to the definition of the energy
flux (heat), the usual area law
of the entropy is not modified but the entropy includes the contribution from
the (modified) gravity.
On the other hand, we may write Eq.~(\ref{BH15}) as
\begin{eqnarray}
\label{BH16}
&&
T^{(\mathrm{matter})}_{\mu\nu}
= \frac{1}{\kappa^2}\left(R_{\mu\nu} - \frac{1}{2}R g_{\mu\nu}\right)
+ G^{\mathrm{modified\ gravity}}_{\mu\nu}\,,
\nonumber \\
&&
G^{\mathrm{modified\ gravity}}_{\mu\nu}
\equiv - T^{(\mathrm{modified\ gravity})}_{\mu\nu}\,.
\end{eqnarray}
If we consider the contribution only from matter to the definition of the
energy flux (heat), in general the entropy $S$ will be expressed by a
function of the area $A$ as $S=h(A)$, where $h(A)$ is an appropriate (not 
always linear) function in terms of $A$ and it may include the parameters 
coming from the modified gravity and/or curvatures, etc.
Furthermore, there might be a mixture of Eqs.~(\ref{BH15}) and (\ref{BH16})
like
\bea
\label{BH17}
&& T^{(\mathrm{matter})}_{\mu\nu} +
\tilde T^{(\mathrm{modified\ gravity})}_{\mu\nu}
= \frac{1}{\kappa^2}\left(R_{\mu\nu} - \frac{1}{2}R g_{\mu\nu}\right)
\nonumber \\
&& \hspace{48mm}
+ \tilde G^{\mathrm{modified\ gravity}}_{\mu\nu}\,, \nn
&& G^{\mathrm{modified\ gravity}}_{\mu\nu}
= - T^{(\mathrm{modified\ gravity})}_{\mu\nu}
\nonumber \\
&& \hspace{0mm}
= \tilde G^{\mathrm{modified\ gravity}}_{\mu\nu}
- \tilde T^{(\mathrm{modified\ gravity})}_{\mu\nu}
\,.
\eea
Thus, the entropy contains the contribution not only from the matter but
from the modified gravity partially, and
the expression of the entropy could be modified from the Einstein gravity.
This may tell that when we discuss the entropy, we may clarify
the contribution to the entropy is purely from the matter or
partially from (modified) gravity.
Then especially in case that the theory includes the scalar field(s), we 
cannot always apply Wald's formula in Eq.~(\ref{BH02}) so naively.


In conclusion, we have explicitly illustrated that the equations of
motion for modified gravity theories, in particular $F(R)$-gravity,
the scalar-Gauss-Bonnet gravity, $F(\mathcal{G})$-gravity and
the non-local gravity, are equivalent to the Clausius relation in
thermodynamics.
In modified gravity theories,
whether we include the contribution from the matter with or without
the modified gravity to the definition of the energy flux (heat) is crucial to
the expression of the entropy.
This point is closely related to the discussion in~\cite{Bamba:2009id}
where it shows that it is possible to obtain a picture of equilibrium
thermodynamics on the apparent horizon in the expanding cosmological
background for a wide class of modified gravity theories
due to a suitable definition of an energy momentum tensor
of the component from modified gravity
that respects to a local energy conservation.


K.B. and C.Q.G. thank Professor Shinji Tsujikawa for very helpful discussions.
K.B. acknowledges the KEK theory exchange program
for physicists in Taiwan and the very kind hospitality at
KEK and Nagoya University.
The work is supported in part by
the National Science Council of R.O.C. under
Grant \#s: NSC-95-2112-M-007-059-MY3 and NSC-98-2112-M-007-008-MY3 and
National Tsing Hua University under the Boost Program and Grant \#:
97N2309F1 (K.B. and C.Q.G.);
MEC (Spain) project FIS2006-02842 and AGAUR (Catalonia) 2009SGR-994, by
JSPS Visitor Program (Japan) and by LRSS project
N.2553.2008.2 (S.D.O.);
and
Global COE Program of Nagoya University provided by the Japan Society
for the Promotion of Science, G07 (S.N.).


\end{document}